# Emergence of a Bandgap in Nano-scale Graphite: A Computational and Experimental Study

Sujinda Chaiyachad[a], Trung-Phuc Vo[b,c], Warakorn Jindata[a], Sirisak Singsen[a], Tanachat Eknapakul[a,d], Chutchawan Jaisuk[a], Patrick Le Fevre[e], Francois Bertran[e], Donghui Lu[f], Yaobo Huang[g], Hideki Nakajima[h], Watchara Liewrian[i], Ittipon Fongkaew[a], Ján Minár[b,*] and Worawat Meevasana[a,*]

[a]School of Physics and Center of Excellence on Advanced Functional Materials, Suranaree University of Technology, Nakhon Ratchasima 30000, Thailand

[b]New Technologies-Research Center, University of West Bohemia, 30100, Pilsen, Czech Republic

[c]Institute of Physics, Czech Academy of Sciences, Cukrovarnická 10, 16200 Praha 6, Czech Republic

[d]Functional Materials and Nanotechnology Center of Excellence, School of Science, Walailak University, Nakhon Si Thammarat, 80160, Thailand

[e]Synchrotron SOLEIL, L'Orme des Merisiers, Départementale 128, F-91190 Saint-Aubin, France

[f]Stanford Synchrotron Radiation Lightsource, SLAC National Accelerator Laboratory, 2575 Sand Hill Road, Menlo Park, California 94025, USA

[g]Shanghai Advanced Research Institute, Chinese Academy of Sciences No.99 Haike Road, Zhangjiang Hi-Tech Park, Pudong Shanghai, P R China

[h]Synchrotron Light Research Institute, Nakhon Ratchasima 30000, Thailand

[i]Department of Physics, King Mongkut's University of Technology Thonburi, Bangkok 10140, Thailand

\* Corresponding author.
*E-mail address:* jminar@ntc.zcu.cz and worawat@g.sut.ac.th

**Abstract**

Bandgaps in layered materials are critical for enabling functionalities such as tunable photodetection, efficient energy conversion, and nonlinear optical responses, which are essential for next-generation photonic and quantum devices. Gap engineering could form heterostructures with complementary materials like transition metal dichalcogenides or perovskites for multifunctional devices. Graphite, conventionally regarded as a gapless material, exhibits a bandgap of ~100 meV in nano-scale patterned highly oriented pyrolytic graphite (HOPG), as revealed by angle-resolved photoemission spectroscopy (ARPES) and Raman measurements. Our state-of-the-art calculations, incorporating photoemission matrix element effects, predict this bandgap with remarkable accuracy and attribute it to mechanical distortions introduced during patterning. This work bridges theory and experiment, providing the direct evidence of a tunable bandgap in HOPG. Beyond its fundamental significance, this finding opens new possibilities for designing materials with tailored electronic properties, enabling advancements in terahertz devices and optoelectronics.

# 1. Introduction

Two-dimensional graphene exhibits remarkable properties with a wide range of applications, including high-mobility metallic transistors [1], magneto-electronic devices [2], quantum computing devices based on quantum electrodynamics [3], and terahertz (THz) optoelectronics [4,5]. For the THz devices, THz radiation can pass through media without ionizing it and exhibits good chemical sensitivity, enabling applications in security screening, medical imaging, and chemical characterization. Graphene has high potential for use in THz optoelectronics due to its surface plasmonic oscillators, which can interact effectively with terahertz frequencies [6,7]. Unlike TMDs, it responds to visible and near-UV frequencies due to the suitability of its energy gap [8]. Graphene also exhibits a photo-carrier multiplication effect, in which a single photon can excite more than one electron through the Auger process [9]. In photovoltaic devices, carrier multiplication can enhance charge carrier density, which could theoretically lead to power conversion efficiencies beyond the 33% limit of single-junction devices. However, graphene light-matter interaction remains weak [10], hindering certain applications. Enhancing the interaction between light and matter in graphene at high-THz resonance frequencies will accelerate the development of THz optoelectronics.

To enhance the interaction between light and graphene, increasing the number of graphene layers can significantly improve light-matter interaction reaching up to 99% in graphite. Graphite plasmonic frequency lies in the THz range (~34 THz), and its surface plasmon polariton resonant characteristics are similar to those of conventional metals [10,11]. Additionally, the narrow bandgap that can be induced in bilayer graphene offers high responsivity in THz devices [12]. It is also well established that light-matter interaction can be enhanced using metasurface arrays with periodic lengths smaller than the wavelength of light [13]. Therefore, it is of interest to investigate whether nanoscale patterning of multilayer graphene with an intrinsic energy gap can further improve the performance of THz optoelectronic devices.

Angle-resolved photoemission spectroscopy (ARPES) is widely used to directly measure the electronic structure and observe the bandgaps of graphene samples [14–16]. ARPES measurements reveal that pristine graphene is a gapless material, with its valence and conduction bands intersecting at the K point of the Brillouin zone, known as the Dirac point [17]. Several approaches have been developed to induce an intrinsic bandgap in graphene, such as substrate-induced effects [14] and quantum size effects in graphene quantum dots (GQDs) or graphene nanoribbons (GNRs), where the bandgap is size-dependent [19]. Here, by using ARPES measurement, we show the bandgap opening of nano highly oriented pyrolytic graphite (nano-HOPG) which is not observed in pristine HOPG. We further investigated the origin of bandgap opening with Raman spectroscopy and density functional theory (DFT) calculations.

## 2. Materials and methods

The square nano-scale patterns of HOPG (nano-HOPG) were prepared on a single crystal HOPG substrate (NT-MDT Spectrum Instruments, HOPG YZA Quality, piece size 7x7 mm., thickness 1.8 mm). Their square patterns were created by using the focused ion beam technique (FE-SEM, Zeiss AURIGA), in which a Ga ion was an ion source, with a source voltage of 300 kV and a source current of 100 pA. The 30-nm beam spot size is sputtered on the substrate at a depth of 15 nm. Their electronic structure was measured using ARPES, performed at beamline 09U1 of the Shanghai Synchrotron Radiation Facility, China, using photon energy 95 eV and a Scienta DA30L electron analyzer. Note that the surface of nano-HOPG was cleaned by heating the sample at 500 ºC before ARPES measurement. The ARPES measurement was performed immediately after cleaning the surface in an ultra-high vacuum at a pressure better than $4 \times 10^{-11}$ Torr. The energy resolution was set at 15 meV, and the sample temperature was maintained at 15 K throughout the experiment. Since the dimension of nano-HOPG pattern is $100 \times 100$ μm$^2$, the smaller beam spot size of $20 \times 30$ μm$^2$ was used to measure the areas of HOPG samples covering on and off-grids nano-scale patterns. To locate the nano-scale pattern, we identified the edges of the HOPG substrate as referenced points. Next, we performed ARPES measurements in the $300 \times 300$ μm$^2$ array with a step size of 100 μm in *x* and *y* directions (i.e., $3 \times 3$ grid). The symmetrized EDCs of band dispersion of the ARPES measurements array is shown in Fig. S1. We also measured the ARPES data at the location far away from the nano-scale pattern at approximately 1 mm to compare and contrast the spectra. In addition, the Raman spectroscopy was used to observe disorder structure at room temperature with a laser wavelength of 532 nm, a beam size of 4 μm, and a laser power of 50 mW.

To cope with spectroscopic and many-body aspects in ARPES, the Green's function Korringa-Kohn-Rostoker (KKR) method [20] was applied under multiple scattering formalism implemented in the SPRKKR package [21]. As a result, corresponding calculations were carried out within the fully relativistic Dirac formalism and spin density functional theory, taking into account spin-orbit coupling (SOC) impacts. No shape approximations are made for the description of the potential as we use the full potential approach. In addition, the exchange-correlation energy functional was treated by local density approximation (LDA) and the ground state computations were conducted using the observed lattice constant [22] of a = 2.456 Å, b = 2.456 Å, and c = 6.696 Å. The maximum value of the angular momentum quantum number is truncated at $l_{max} = 3$. Lloyd's formula has been employed for accurate determination of the Fermi level [23]. To achieve convergence in the multiple scattering between layers, our calculations utilized a plane-wave basis, increasing the surface reciprocal lattice vectors $\vec{G}$ to 37. A key parameter in describing multiple scattering involves the expansion of all physical quantities using $l_{\max}$, employing Bauer's identity to represent plane waves (interlayer scattering) with spherical waves (intralayer scattering). The final state is represented by a time-reversed LEED (TR-LEED) state from Pendry's model [24], allowing the accurate description of these states in a wide range of photon energies (from 6 eV up to several keV). Strocov et al. demonstrated that TR-LEED treatment can disentangle the multiband final states giving rise to the broadening of the ARPES peaks in the Ag metal. For obtaining as close as agreements with experiments, ARPES calculations were conducted based on the one-step model (1SM) of photoemission [25,26]. Consequently, the theory accounts for influences triggered by the light polarization, matrix-element effects and surface effects [27,28]. Recently, 1SM simulations were successfully in capturing the typical characteristics of a new type of magnetism, so-called alter-magnetism [29]. This theoretical work provides significant benefits for the emerging field of spintronics, a next-generation magnetic memory technology. The solid surface in the current study is described by semi-infinite lattices with ideal lateral translation invariance and arbitrary number of atoms per unit cell. The computational model described here is applied to the valence band in this work but is also effective for core levels [30-31]. More details regarding the core-level implementation can be found in Ref [32], which elucidates the photoelectron effects responsible for the pronounced modulation of ARPES patterns. To further understand the strain dependence, we use the density functional theory (DFT) calculation as implemented in the VASP code [33-34] to calculate the band dispersion along $k_z$ direction of the 1.1% for biaxial tensile strains. The generalized gradient approximation (GGA) of Perdew-Burke-Ernzerhof (PBE) was used to describe the exchange-correlation functional [35].

The electron-ion interactions were treated by the projector-augmented-wave (PAW) method [36]. Van der Waals correction of the Grimme method (DFT- D3) has been taken into account to explain long-range interactions [37]. The cutoff energy of 520 eV was set to represent number of plane-wave basis sets. The Monkhorst-pack scheme [38] of 21×21×1 k-point mesh for Brillouin interaction was used for the calculation of structural optimization and band structure. The structure was fully relaxed until the atomic force acting on each atom was less than 0.01 eV/Å. The band structure calculation of the structural trilayer graphene was comprehensively regarded to the suggestion in the previous works to obtain an accuracy [39].

## 3. Results and discussion

Figures 1a and 1b show SEM images of nano-HOPG fabricated on a single-crystal HOPG substrate. A schematic illustration of the square-shaped nano-HOPG array, covering an area of 100 × 100 μm² on the substrate, is presented in Fig. 1c. In this work, we aimed to fabricate HOPG squares as small as possible, achieving a minimum size of 300 nm with our current setup. Notably, the nano-HOPG array retains the single-crystal nature of the original HOPG substrate. Figure 1d displays the Fermi surface map of nano-HOPG, which appears circular (red dashed line) around the Brillouin zone (blue solid line). Although HOPG exhibits azimuthal disorder typical of graphite, the band dispersion remains invariant with respect to the azimuthal angle [40]. Figure 1e shows the valence band dispersion of HOPG measured along the direction indicated by the black arrow in Fig. 1d. To investigate bandgap opening, we focus on the Dirac cone near the Fermi level, as highlighted in the inset of Fig. 1e. Figure 2q illustrates the ARPES measurement positions at three distinct regions: (i) the nano-HOPG pattern (A), (ii) the region near the nano-HOPG pattern (B), and (iii) the HOPG substrate (C). The EDCs of the valence band dispersion for nano-HOPG and the HOPG substrate, shown in Fig. 2r, reveal distinct features near the Fermi level. A bandgap opening is observed in the nano-HOPG pattern, as indicated by the symmetrized EDCs in the ARPES spectrum presented in Fig. 1f.

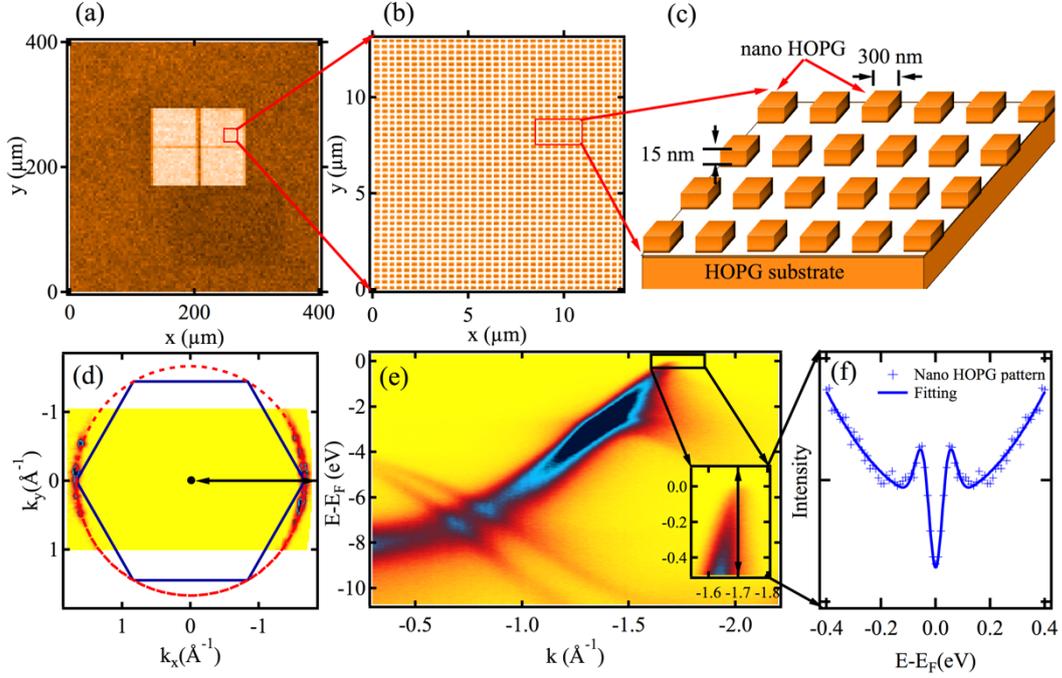

**Fig. 1.** (a) and (b) SEM image of nano-scale HOPG. (c) Schematic of nano-scale HOPG on HOPG substrate. Electronic structure of highly oriented pyrolytic graphite (HOPG) (measured by using 60 eV of photon energy) consist of (d) Femi surface map of HOPG and (e) Band dispersion along black arrow in fig. (d). Inset fig. in (e) is band dispersion near Fermi level in black solid box. (f) Symmetrized EDC spectrum across $k_F$ of nano-HOPG (measured by using 95 eV of photon energy).

Figure 2 presents the electronic structure measured using a photon energy of 95 eV. Figures 2a–2d show the band dispersion near the Fermi level for nano-HOPG, the HOPG substrate, the region near nano-HOPG, and gold (Au), respectively. These correspond to the energy dispersion curve (EDC) spectra at $k_F$, shown in Figures 2i–2l. Typically, graphene or graphite exhibits a Dirac cone at the K point of the Brillouin zone near the Fermi level. In the case of nano-HOPG, we observe a density of states peak in the EDC at approximately 59 meV below the Fermi level, indicating the presence of a bandgap. This bandgap opening in the nano-HOPG pattern is clearly visible in the symmetrized ARPES spectra, using the Fermi level of Au as a reference. The symmetrized band dispersions near the Fermi level for nano-HOPG, the HOPG substrate, the region near nano-HOPG, and Au are shown in Figs. 2e–2h, respectively. The corresponding symmetrized EDC spectra are presented in Figures 2m–2p. A bandgap of 112 ± 15 meV is observed in the symmetrized EDC spectrum of nano-HOPG (Fig. 2m). The curves in Figs. 2m–2p are included as visual guides. The ARPES data of the nano-HOPG was measured at the beamline 5-4, Stanford Synchrotron Radiation Lightsource (USA) for a reproducibility test, which shows a similar result. The bandgap opening was reproducibly observed in nano-HOPG shown in Fig. S2.

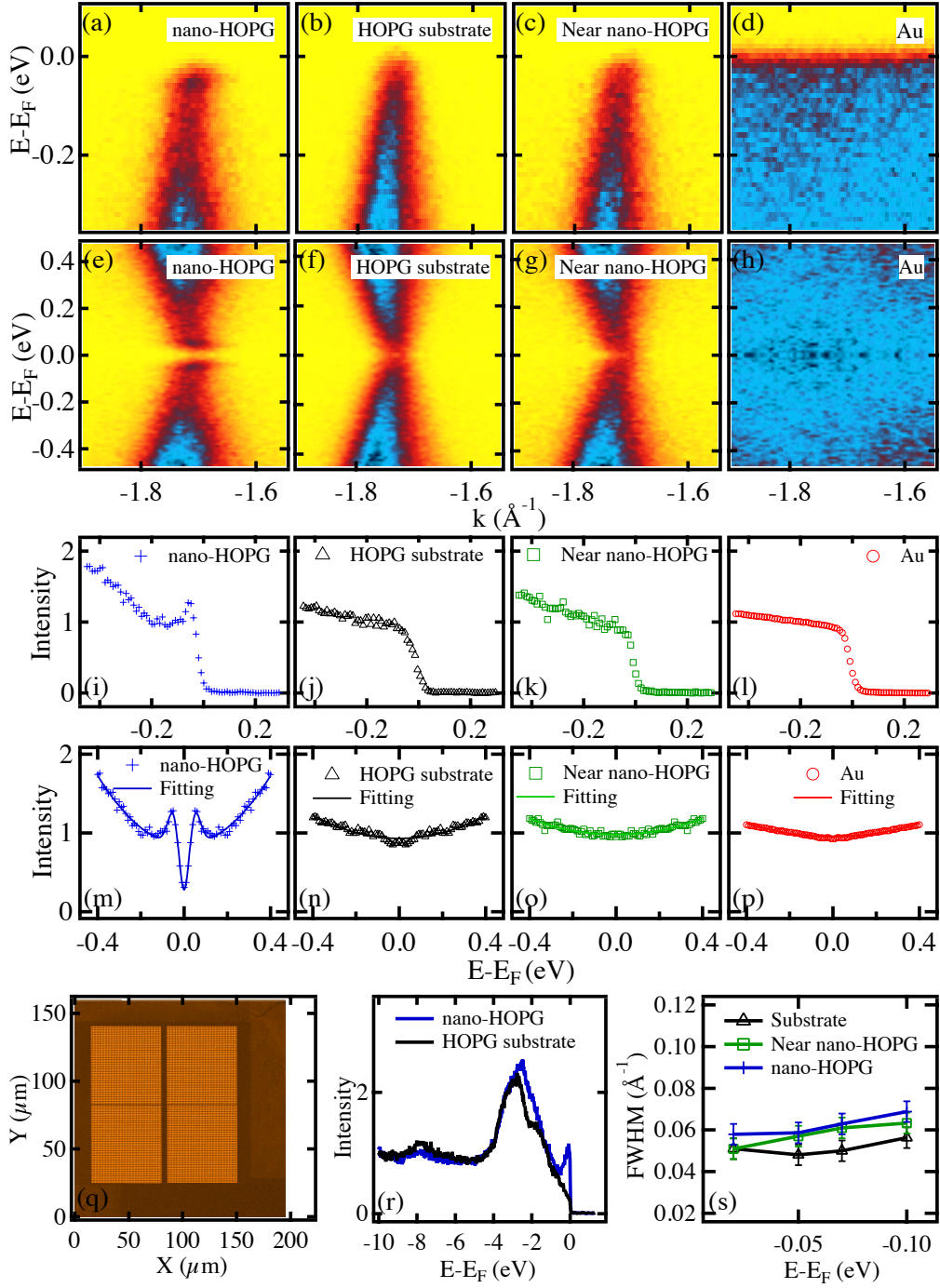

**Fig. 2.** The ARPES were measured by using 95 eV of photon energy. (a) – (d) ARPES spectra of nano HOPG pattern, HOPG substrate, near nano HOPG pattern, and gold at near Fermi level, respectively. (e) – (h) Symmetrized ARPES spectra of (a) – (d) respectively. (i) – (l) EDC spectra across $k_F$ of ARPES spectra (a) – (d) respectively. (m)-(p) EDC spectra across $k_F$ of symmetrized ARPES spectra (e) – (h). (q) ARPES

measurement positions of nano HOPG pattern (A), near nano HOPG pattern (B) and HOPG substrate (C) that are far from the pattern by 1 mm. (r) EDC spectra across $k_F$ of the valence band of the nano HOPG pattern and the HOPG substrate. (s) FWHM of MDC spectra of nano HOPG pattern, near nano HOPG pattern and HOPG substrate.

Here, this work showed different explanation on bandgap opening in the HOPG graphene-family by using FIB pattering. This is counterintuitive to the predicted quantum confinement. In the later discussion, we try to explain why it is not from quantum confinement rather it is likely to form by tensile strain. Achieving band gap opening in HOPG is crucial or comparable to what observed in graphene. Regarding the origin of the bandgap opening, quantum confinement is a primary candidate due to the nanoscale dimensions of the sample. Typically, nanosized graphite-based materials can exhibit semiconducting behavior, with a bandgap arising from quantum confinement effects. The shrinking of the bandgap depends on the size of the nanoparticle. This effect is dominant in the electronic structure of a few-nanometer particle [41]. In this work, the energy bandgap of the nano HOPG pattern is larger than other graphene-base nanoscale (GQD, GNR) at the same size, up to 1 order of magnitude [21]. Therefore, quantum confinement is unlikely to be the primary cause of the observed bandgap in our case.

We then explored other possible causes for the observed bandgap opening. One possibility considered was the twist angle of the top layer in nano-HOPG, given that HOPG is azimuthally disordered. In multilayer graphene, specific twist angles can result in flat bands that may contribute to bandgap formation [1]. However, in our measurements, no flat band was observed in the nano-HOPG where the bandgap appeared. This suggests that the bandgap is not caused by a twist angle between graphene layers. Regarding to the studies of ion beam effect on sample surface, the ion beam created edge of nano hole or nano island that dominated by zigzag edge.

Moreover, the ion beam induces disordered structures, predominantly composed of sp³ hybridized carbon, as observed through Raman spectroscopy. The degree of disorder depends on the ion dose; at high ion doses, the HOPG surface can become amorphous [42–43]. Notably, sp³-type defects have been shown to induce a transition from metallic to semiconducting behavior in multilayer graphene [44]. Figure 3 presents the Raman spectra of the HOPG substrate, the region near the nano-HOPG pattern, and the nano-HOPG pattern. The disorder peak (D peak) of the nano HOPG pattern and near nano-HOPG pattern, shown in green and blue solid lines, reveals an ion beam sputtered disorder defect while the D peak is not present in the normal HOPG surface, shown in black solid line. To further examine the effect of disorder, we analyzed the full width at half maximum (FWHM) of the momentum distribution curves (MDCs) extracted from the ARPES data [45]. As shown in Figure 2s, the FWHMs for both the nano-HOPG and near-nano-HOPG regions are comparable. Since the bandgap opening is only observed in the nano-HOPG pattern, we conclude that disorder is not the primary cause of the bandgap opening.

We consider the effect of strain on the nano-HOPG. Strain influences the vibrational frequencies of molecular bonds, which can be detected as shifts in Raman spectral features. Both the G and 2D modes in the Raman spectrum correspond to in-plane C–C vibrational modes. These modes shift to higher frequencies under compressive strain or due to oxidation, as observed in graphene oxide and oxidized graphite [46]. Conversely, tensile strain causes a shift to lower frequencies. Due to the broad spectra of G mode, 2D mode was used to be characterized. 2D mode of Raman spectrum was divided into two Lorentzian peaks. By Lorentzian fitting, the peak position of second peak of the nano-HOPG is significantly redshifted which is 4 cm$^{-1}$ compared to the HOPG surface as shown in inset of Fig. 3. The redshift refers to a decrease in frequency of 2D mode which may confirm that the organic bonding is not dominated on nano-HOPG. Moreover, there is a possibility that this redshift may come from expansion of carbon hexagonal ring which is called tensile strain [47]. Similar strain effects have been reported in nanopatterned monolayer $MoS_2$, where strain modulates photoluminescence properties [48]. Based on these observations, we suggest that a small amount of tensile strain may be present in the nano-HOPG.

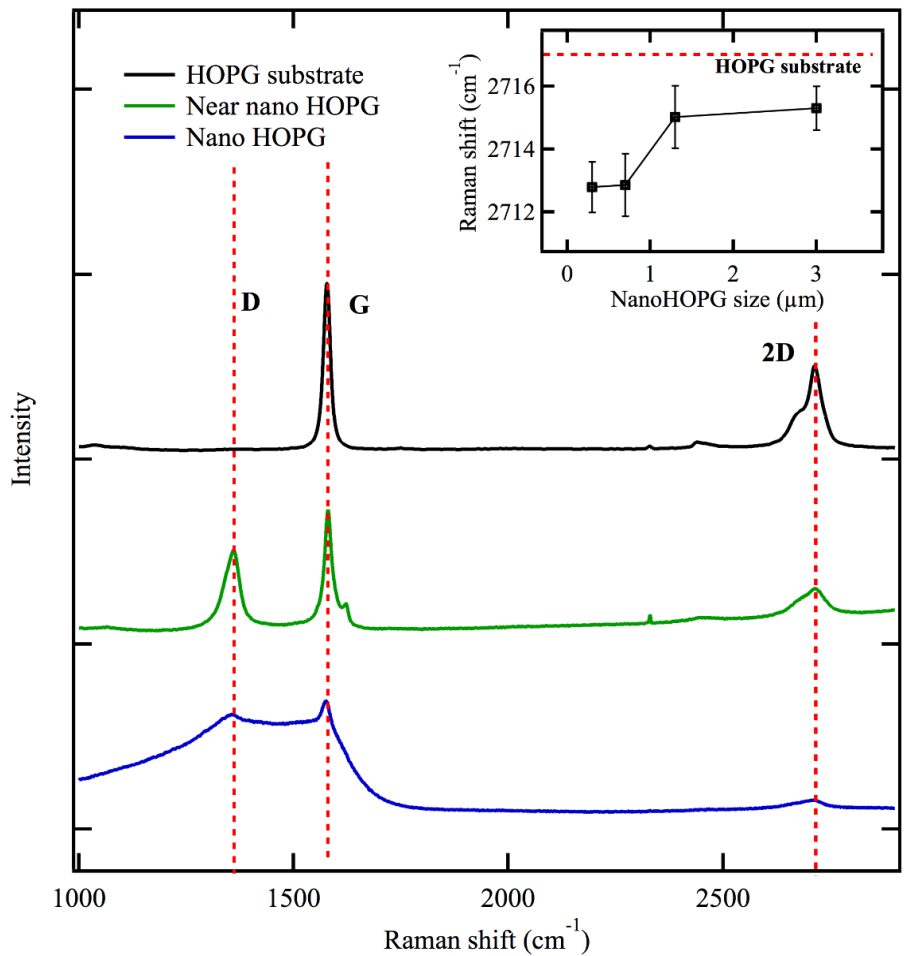

**Fig. 3.** Raman spectra of HOPG substrate, near nano-HOPG pattern and nano-HOPG pattern. The peak positions of the D band, G band, and 2D band were indicated by the red dashed lines. The inset shows the fitting of peak positions of the 2D bands of each nano-HOPG size and HOPG substrate.

To obtain insight in the strain effect on the bandgap opening observed in nano-HOPG, we performed the electronic structure calculations under strain effect shown in Fig. 4. For illustrative purposes, the unit cell of the distored and pristine HOPG are superimposed (Fig. 4a), implying the structural difference between a pristine and distorted case with AB stacking sequence. In this configuration, the second C sub-layer of the first layer (from top) contains distortion. To investigate the distortion dependence of HOPG electronic structure, one C atom is shifted close to the other along $\vec{b}$ direction by distorted values $\Delta(\text{Å}) \in [-0.155, -0.006]$ with step 0.025 Å. Fig. 4b illustrates the selected band dispersion by compressing $\Delta(\text{Å}) = [-0.155, -0.105, -0.006]$ and one case of stretching $\Delta(\text{Å}) = 0.042$ along $\Gamma - K$, computed by one-step photoemission theory, of various HOPG configurations. The position of the Fermi level was rearranged for the DFT calculations to align the top of the calculated valance band with the one from ARPES measurements, namely adding 0.08 eV to the binding energy as an offset. The position of high-symmetry points differs to literature due to impacts of the incident angle of light. Clearly, highly dispersive bands are found along this direction. For instance, there is always the fact that one band possesses the top of dispersion at the $\Gamma$ point and disperses toward the higher binding energy on approaching the $K$ point. Deformation brings several trends regarding to distorting level. When $|\Delta|$ reduces from 0.155 Å to 0.006 Å, some top-most bands around $\Gamma$ move down and overlay to others. In contrast, those bands surrounding $\Gamma$ starts to split and go up as $\Delta$ reaches 0.042 Å. Furthermore, the band in the vicinity of $K$ is shifted to the right side. Noticeably, at the beginning the energy bandgap emerges, reduces and closes later on. Figs. 4c and 4d show the simulated stretched and raw-data experimental spectra of investigated HOPGs. Calculated models take into consideration the experiment geometry and as many parameters as possible that were used in experiments. From computational point of view, stretching the topmost layer by 0.080 Å along $\vec{b}$ direction creates a bandgap $E_g$ = 137 meV meanwhile $E_g$ = 112 meV is observed in ARPES measurements. In general, there a is good agreement between computational and measured data in terms of band-gap appearance and major dispersion features around high-symmetry points. Nevertheless, we notice a difference concerning the band-gap value. Partially the later discrepancy might be due to the low accuracy of our DFT exchange-correlation functional. Overestimation can happen when considered band gaps are smaller than 1 eV [49]. HOPG is well-known as a semimetallic material which is certified by many previous studies [50]. However, it can behave semiconducting by being irradiated by visible polarized light perpendicular to its planes [51]. By various techniques (e.g., scanning tunneling microscope (STM) or temperature-dependent resistance), energy gap of graphite is found in order of ~30 – 40 eV [52,53]. The best HOPG samples are supposed to own crystallites preferentially oriented in such a way that mosaic angle spread is less than one degree [54]. Consequently, electronic properties strongly depend on the alignment of graphite crystals along c-axis. Depending on how layers are stacked, HOPG is classified into either well-ordered structures (metallic) or poorly ordered ones (semiconducting) [55]. In addition, stacking fault defects are proposed to be a cause of anomalously high resistivity and quantum transport in HOPG [56,57]. Experimentally, horizontal shifts in the HOPG top layer, produced by a STM tip, lead to transitions between ABA and ABC stacking [58]. Therefore, not surprisingly our work indicates the semiconductor fingerprint and one potential trigger for the bandgap creation of HOPGs, which is in harmony with above-mentioned papers. Our introduced energy bandgap (137 meV and 112 meV) are greater than former reports [59-60]. It is noteworthy that this inequality is likely from different scanned sample areas, energy resolution and not mentioned parameters in ARPES publications. To further confirm the bandgap opening at *K* point, we calculated the photon energy dependence of ARPES of the stretched configuration displayed in Fig. 4c. Energy distribution curves (EDCs) imply that constituting states are gaped within the photon energy range differently (Fig. 4d). Thus, our findings emphasize the confidence interpretation of a bandgap existing in HOPGs.

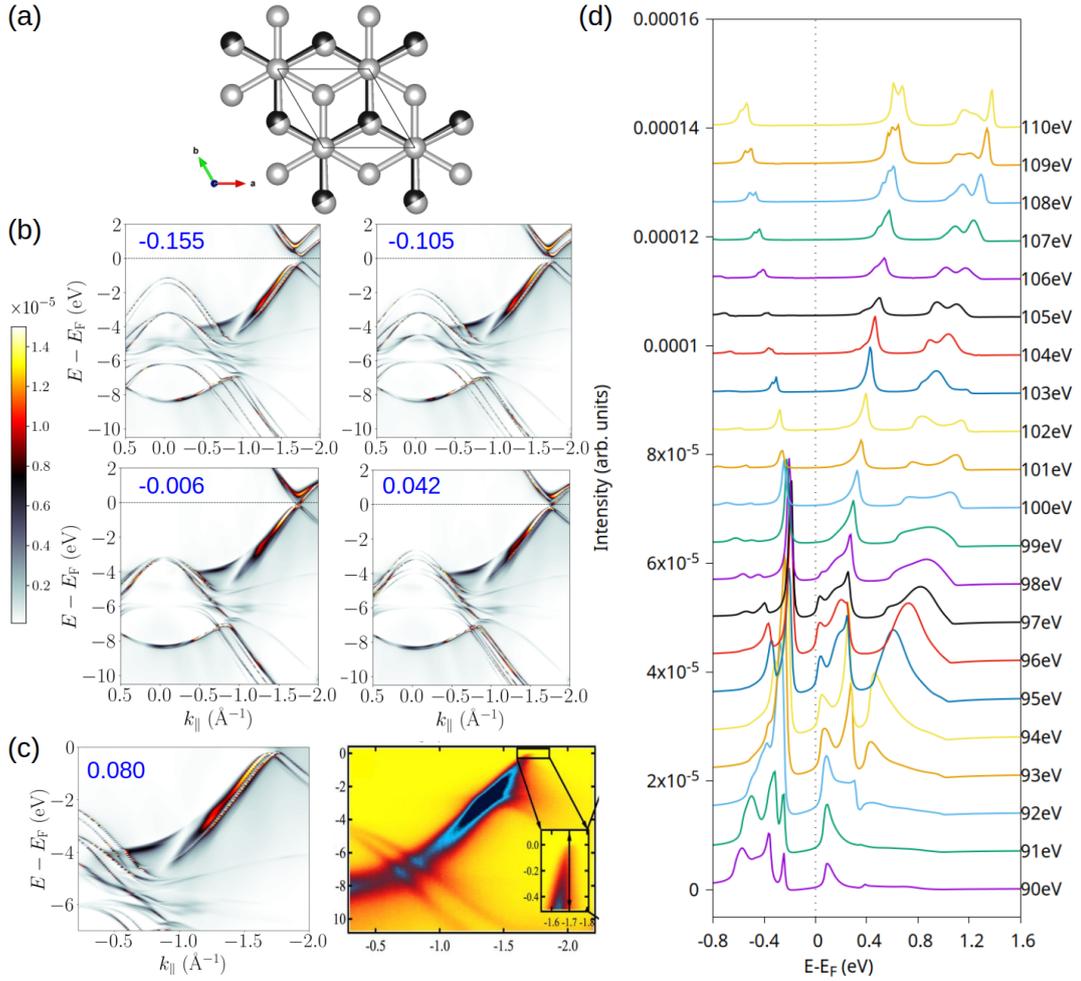

**Fig. 4.** Electronic band structure of bulk HOPG. (a) Top view of the distorted (gray, top) and pristine (black, bottom) HOPG unit cells, shown superimposed for illustrative purposes. (b) ARPES intensity map calculated at a 95 eV photon energy along the $\bar{\Gamma}$ - $\bar{K}$ line under selectively compressing (0.155, −0.105, −0.006) and stretching

(0.042 Å) structures along (as illustrated in (a)). The blue labels indicate distorting values Δ. (c) Similar to (a) but with Δ = 0.080 Å. (d) Energy distribution curve (EDC) computed at $K$ point as a function of photon energy.

In addition to the above-mentioned multiple scattering method, we carried out the electronic band structure calculations by the pseudopotenial method. Herein the investigation is applied a biaxial tensile strain. The relaxed structural unit cell of trilayer graphene was used for simulation multi-layer effect of graphene in HOPG, as shown in Fig. 5(a). In order to study band-gap opening, we considered the electronic band structure of trilayer graphene compared with monolayer graphene in the presence of biaxial tensile strain varying from 0 to 2% according to the schematic as shown in Fig. 5(b). From the experimental result, the Raman spectra of the 2D peak are shifted to lower frequency relating to the increase of the C-C bond length. In addition, the Raman 2D band corresponds with the iTO (A1) phonon modes in D3h symmetry, in which all carbon atoms on the carbon ring move towards and move outward the Γ-center. Therefore, the biaxial tensile is a possible type of strain that corresponds with 2D band vibration mode and plays an important role in expanding C-C bond length in all directions. The comparative energy gap as a function of biaxial tensile strain is depicted in Fig. 5(c). Note that the calculated energy gaps of graphene with applied strain are lower than the experimental values due to the common self-interaction error of DFT-PBE calculation [61]. The finding shows that the energy gap of monolayer graphene tends to linearly increase, whereas that of trilayer graphene is roughly increasing. The band structure of trilayer graphene in the absence of strain is shown in Fig. 5(e). During applied 0 to 2% strain, computational investigation reveals that the highest gap opening of trilayer graphene is under 1.1% strain with an energy gap of 93.0 meV, which is close to the measured energy gap of HOPG (112 meV). Meanwhile, that of monolayer graphene is equal to 80.9 meV, which is lower. The corresponding band structure at K-point is illustrated in Figs. 5(d) and 5(f) for monolayer and trilayer graphene at 1.1% strain, respectively. According to this computational result, it suggests that the biaxial tensile strain of trilayer graphene is likely responsible for the gap opening of HOPG.

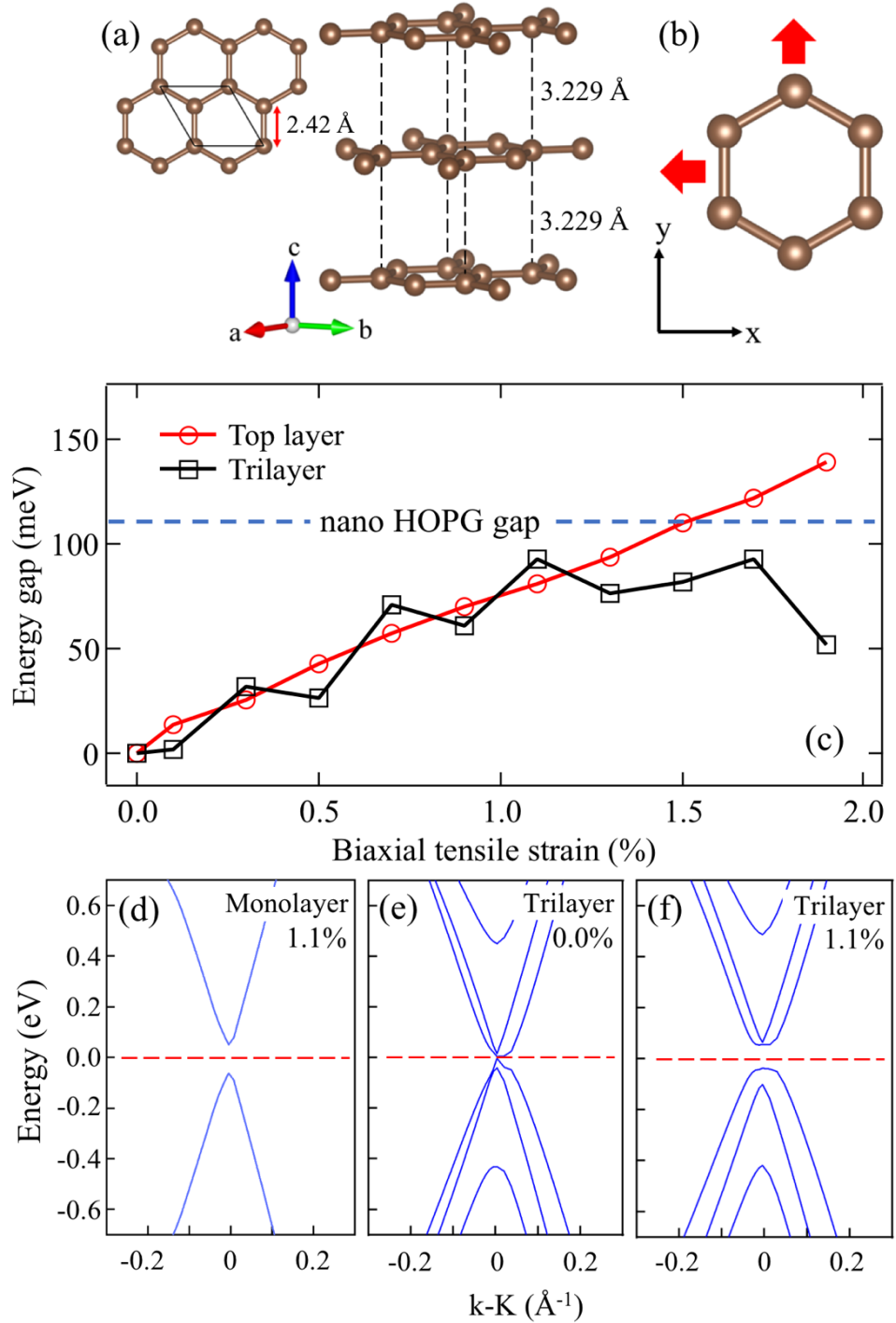

**Fig. 5.** (a) The geometric structure of trilayer graphene with corresponding C-C band distance and the interlayer spacing. (b) Schematic of biaxial tensile strain with a same strain value along x- and y-direction. (c) The comparative energy gap of monolayer, trilayer graphene, and nano HOPG. The corresponding band structure of monolayer

graphene with applied 1.1% strain (d), trilayer graphene with no strain (e), and trilayer graphene with applied 1.1% strain (f).

Moreover, we use the DFT calculation to study the energy band dispersion along $k_z$ direction of the 1.1% biaxial tensile strain on three layers of graphene as shown in Fig. 6. We observed that the dispersion clearly shows no bandgap closing at all k points. From the ARPES data at different $k_z$ (see Fig. S3) and the DFT calculations, it suggests to the observed bandgap opening through the Brillouin zone along $k_z$. Note that the same behavior along $k_z$ is also observed in the similar system $MoS_2$ [62].

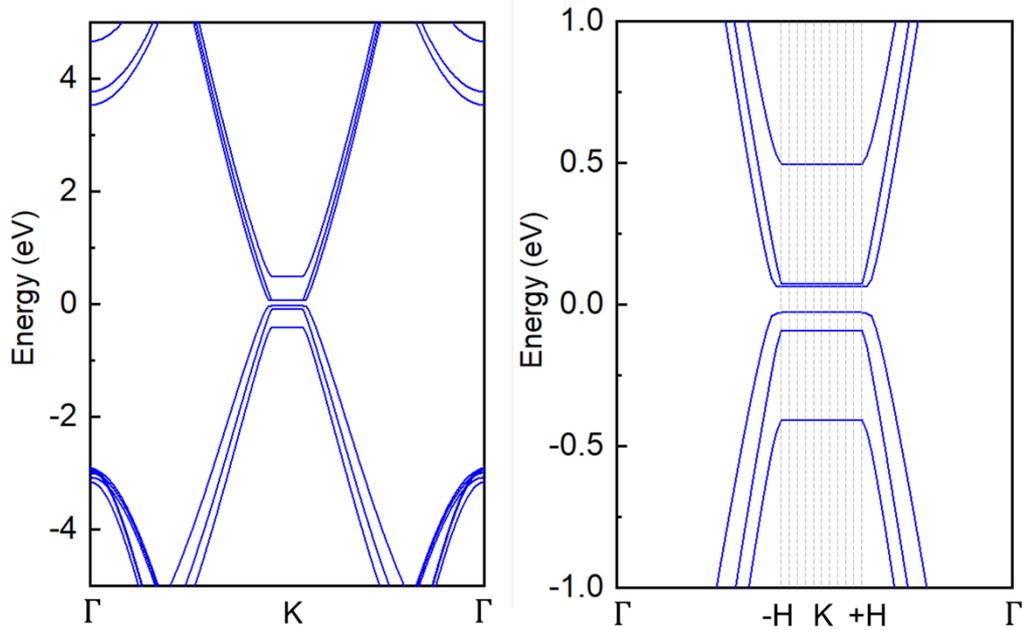

**Fig. 6.** Band structure of three layers graphene with 1.1 % of biaxial tensile strain.

We have shown the bandgap opening of nanoscale crystalline exists in the hundreds of meV range, which corresponds to the energy range from mid-infrared to terahertz. In comparison to previous research that found bandgaps in single crystal graphene, the bandgaps created by the various methods are identical in the hundreds of meV range. For example, an electric field generated in bilayer graphene [16], whereas graphene grown on a SiC substrate has a bandgap of 260 meV for a single layer and decreases with the number of layers [15]. Additionally, it is effective when the tensile strain of graphene with a bandgap of 300 meV at 1% tensile strain is calculated, and the energy bandgap reduces as the number of layers rises with the same tensile strain [31]. As a result, the bandgap of nanoscale crystalline determined by theoretically and experimentally in this study is comparable to that of graphene crystalline determined in previous research investigations.

The ability to open a bandgap in graphene and related materials has been a long-standing goal due to its critical role in enabling semiconducting and optoelectronic applications. In monolayer or bilayer graphene, various approaches such as lateral quantum confinement in nanoribbons [63], patterned epitaxial structures [64], or periodic adsorption patterns like hydrogen superlattices [65] have demonstrated bandgap openings typically attributed to quantum confinement or symmetry-breaking mechanisms. These studies laid the foundation for bandgap engineering in 2D carbon systems, yet most remain constrained to single- or few-layer graphene with controlled environments. In contrast, the present work demonstrates that even in multilayered, quasi-bulk systems such as HOPG, a bandgap of ~100 meV can be achieved through nanoscale patterning. This is significant not only because it expands bandgap engineering to thicker graphitic materials but also because the underlying mechanism appears to diverge from previously understood quantum confinement effects. Using focused ion beam (FIB) patterning, we created nanoscale square arrays on the HOPG surface and observed a well-defined bandgap via ARPES, corroborated by Raman spectroscopy and DFT calculations. Importantly, the electronic structure does not reflect characteristics of lateral confinement; instead, our analysis indicates that tensile strain introduced by surface reconstruction is the most plausible origin of the observed bandgap. This interpretation is supported by theoretical models for strained configurations of semi-infinite and trilayer graphite. Moreover, prior studies on finite-sized graphite with Bernal stacking report only narrow intrinsic gaps (~40 meV), and that interface and structural effects may dominate over pure confinement [66]. Thus, our results introduce a novel mechanism strain-induced bandgap opening in nano-HOPG that is structurally driven rather than size-limited, offering a distinct and scalable route for modulating the electronic properties of graphitic materials. This contribution complements and extends previous strategies developed for graphene, underscoring the broader potential of structural patterning in graphite-family materials for next-generation electronic and photonic applications.

## 4. Conclusions

In summary, we show the energy bandgap opening in the order of 100 meV occurred at the surfaces of nano-scale HOPG square patterns by calculations from 2 different methods (full-potential scattering and pseudopotential in conjunction with variational principles), supporting the experiment ARPES observation. The Raman shift and DFT calculations suggest that tensile strain is likely the origin of this bandgap opening with two strained configurations. The semi-infinite crystal with 0.080 Å distortion in the top-most layer possesses an energy bandgap of 137 meV. The trilayer model under the biaxial tensile strain reveals the gap of 93 meV. The present study connects strain effects to photoemission and proposes solutions for band-gap engineering, which opens the door to several promising applications across tunable photodetectors and light-emitting devices. Our finding also demonstrates how to create a crystalline nanopattern whose energy bandgap is compatible with the terahertz application. We believe that our method will also provide a route to modify electronic structure of other crystalline materials/metasurfaces [67] which will be useful not only in the terahertz devices but also other optoelectronics.

**Declaration of competing interest**

The authors declare that they have no known competing financial interests or personal relationships that could have appeared to influence the work reported in this paper.

**Acknowledgments**

This work was supported by the Program Management Unit for Human Resources and Institutional Development, Research and Innovation (Thailand), Grant No. B39G670018, Suranaree University of Technology, Thailand Science Research and Innovation (TSRI) and National Science Research and Innovation Fund (NSRF). S.C. acknowledges DPST for financial support. This work was also supported by the project Quantum materials for applications in sustainable technologies (QM4ST), funded as project No. CZ.02.01.01/00/22 008/0004572 by Programme Johannes Amos Commenius, call Excellent Research (T.-P.V., J.M.) and the Czech Science Foundation Grant No. GA ČR 23-04746S (T.-P.V.). The ARPES measurements are supported by beamline 09U1 of the Shanghai Synchrotron Radiation Facility, the CASSIOPEE beamline, SOLEIL synchrotron, and beamline 5-2, Stanford Synchrotron Radiation Lightsource. We would also like to thank T. Taychatanapat, W. Saengsui, P. Laohana, A. Mooltang, S. Musikajareon, P. Chanprakhon, S. Polin and A. Rasritat for useful discussion and help with measurements.

**Data availability**

Data will be made available on request.

**Supplementary Material**

See the Supporting Material for detail on symmetrized EDC spectra across $k_F$ of each position of ARPES measurements in the 300 x 300 $\mu m^2$ array with a step size of 100 μm in x and y directions (i.e., 3 × 3 grid) and Reproduced of ARPES measurements at Stanford Synchrotron Radiation Lightsource, beamline 5-4. Moreover, we include the data of the observation of gap opening by using different photon energies (i.e., at different kz).

Supplementary Material for

# Emergence of a Bandgap in Nano-scale Graphite: A Computational and Experimental Study

Sujinda Chaiyachad[a], Trung-Phuc Vo[b,c], Warakorn Jindata[a], Sirisak Singsen[a], Tanachat Eknapakul[a,d], , Chutchawan Jaisuk[a], Patrick Le Fevre[e], Francois Bertran[e], Donghui Lu[f], Yaobo Huang[g], Hideki Nakajima[h], Watchara Liewrian[i], Ittipon Fongkaew[a], Ján Minár[b,*] and Worawat Meevasana[a,*]

[a]School of Physics and Center of Excellence on Advanced Functional Materials, Suranaree University of Technology, Nakhon Ratchasima 30000, Thailand
[b]New Technologies-Research Center, University of West Bohemia, 30100, Pilsen, Czech Republic
[c]Institute of Physics, Czech Academy of Sciences, Cukrovarnická 10, 16200 Praha 6, Czech Republic
[d]Functional Materials and Nanotechnology Center of Excellence, School of Science, Walailak University, Nakhon Si Thammarat, 80160, Thailand
[e]Synchrotron SOLEIL, L'Orme des Merisiers, Départementale 128, F-91190 Saint-Aubin, France
[f]Stanford Synchrotron Radiation Lightsource, SLAC National Accelerator Laboratory, 2575 Sand Hill Road, Menlo Park, California 94025, USA
[g]Shanghai Advanced Research Institute, Chinese Academy of Sciences No.99 Haike Road, Zhangjiang Hi-Tech Park, Pudong Shanghai, P R China
[h]Synchrotron Light Research Institute, Nakhon Ratchasima 30000, Thailand
[i]Department of Physics, King Mongkut's University of Technology Thonburi, Bangkok 10140, Thailand

* Corresponding author.
*E-mail address:* jminar@ntc.zcu.cz and worawat@g.sut.ac.th

Figure S1 shows the symmetrized EDC of ARPES spectra across k$_F$ of HOPG surface that were performed at beamline 09U1 of the Shanghai Synchrotron Radiation Facility, China. This ARPES data were measured by using 95 eV of photon energy. This figure clarifies where the nano HOPG is located in the experiment section. Each position of ARPES measurements in the 300 x 300 μm$^2$ array with a step size of 100 μm in x and y directions (i.e., 3 × 3 grid). The nano-HOPGs are located at positions (i), (ii), and (iv).

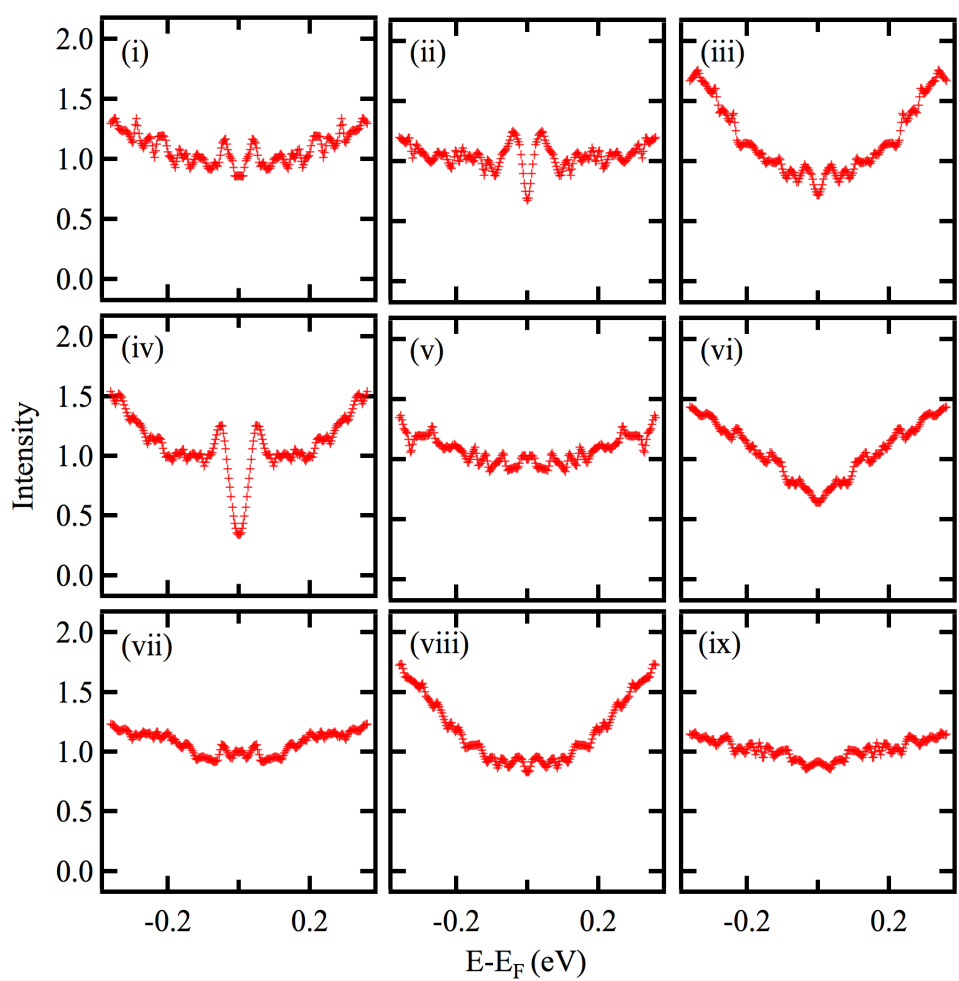

**Fig. S1** (i)-(ix) Symmetrized EDC spectra across k$_F$ of each position of ARPES measurements in the 300 x 300 μm$^2$ array with a step size of 100 μm in x and y directions (i.e., 3 × 3 grid).

Figure S2 illustrate the symmetrized EDC of ARPES spectra across k$_F$ of nano-HOPG(i), HOPG surface(ii), and Au(iii). This data was measured by using 60 eV of photon energy at the beamline 5-4, Stanford Synchrotron Radiation Lightsource, in the USA.

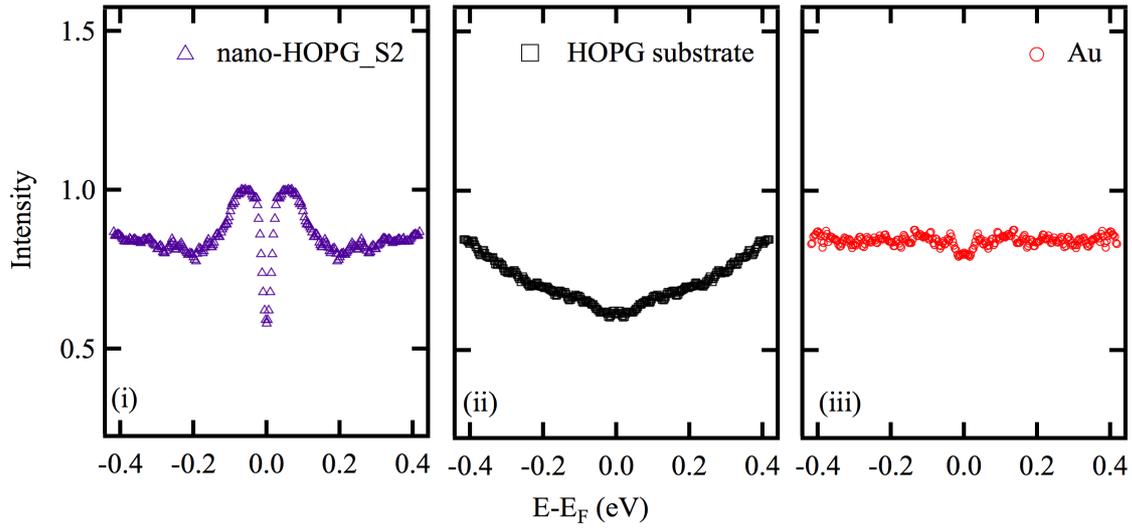

**Fig. S2** ARPES measurements at Stanford Synchrotron Radiation Lightsource, beamline 5-4, (i)-(iii) symmetrized EDC spectra across k$_F$ of ARPES spectra of nano-HOPG, HOPG substrate, and Au respectively.

From the ARPES measurement, we have observed the gap opening using different photon energies (i.e., at different k$_z$) as shown in Fig. S3 below 60 eV (Figs. S3a–S3d) and 95 eV (Figs. S3e–S3h).

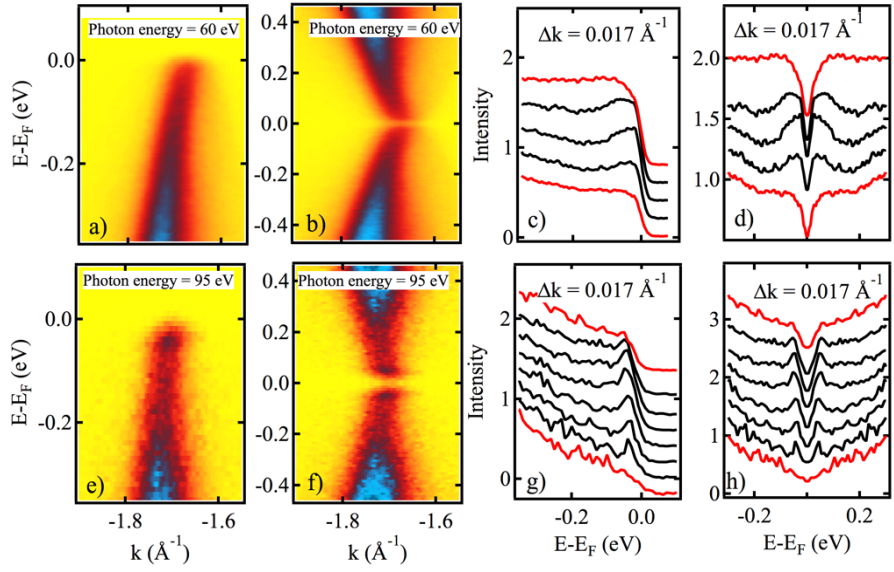

**Fig. S3** ARPES spectra of nano HOPG that was measured at 60 eV (a) and 95 eV (e). The symmetrized ARPES spectra of (a) and (e) are shown in (b) and (f), respectively. EDC spectra near $k_F$ of (a) and (e) in (c) and (g), respectively. EDC spectra near $k_F$ of (b) and (f) are shown in (d) and (h), respectively.

4